\documentclass[reprint,groupedaddress,amsmath,amssymb,pra]{revtex4-2}

\usepackage{hyperref}
\usepackage{amsmath} 
\usepackage{amssymb}
\usepackage{physics}
\usepackage[pdftex]{graphicx}
\usepackage{cleveref}

\usepackage{setspace}
\usepackage{titlesec}

\usepackage{color}
\usepackage[shortlabels]{enumitem}

\usepackage[normalem]{ulem}

\usepackage{booktabs}

\newcommand{\mrm}{\mathrm}

\newcommand{\avg}[1]
    {\left\langle {#1} 
    \right\rangle }

\newcommand{\Esca}{\mathcal{E}}

\newcommand{\Bsca}{\mathcal{B}}

\begin{document}

\title{Coherent quantum beats: spectroscopy of energy differences masked by inhomogeneous broadening}

\author{H. D. Ramachandran}
\author{J. E. Ford}
\author{A. C. Vutha}
\affiliation{Department of Physics, University of Toronto, Toronto ON M5S 1A7, Canada}

\begin{abstract}
Precision spectroscopy of solid-state systems is challenging due to inhomogeneous broadening. We describe a technique -- coherent quantum beats -- that enables the measurement of small frequency shifts within an inhomogeneously broadened distribution while addressing the full ensemble. We show that the technique can be used to obtain improvements in signal size and spectral resolution, offering advantages for precision measurements in solids.
\end{abstract}

\maketitle

\begin{figure}
    \centering
    \includegraphics[width=0.75\columnwidth]{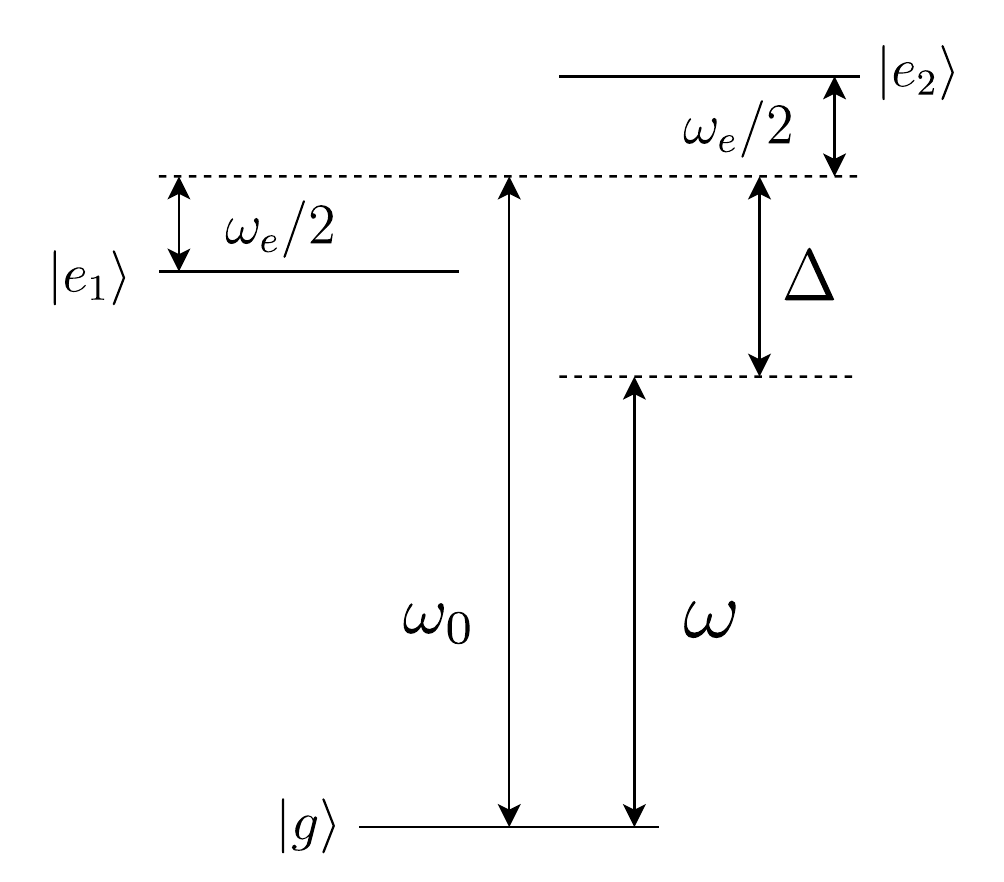}
    \caption{A system of three levels, where the energy difference of interest ($\omega_e$) is measured using spectroscopy of the $g \to e_1$ and $g \to e_2$ transitions. The frequency of light is $\omega$ and the average of the $g \to e_1,e_2$ resonance frequencies is $\omega_0$. The driving frequency is detuned from the resonant frequency by $\Delta$.} 
    \label{fig:3_level_structure}
\end{figure}

\begin{figure}
    \centering
    \includegraphics[width=0.8\columnwidth]{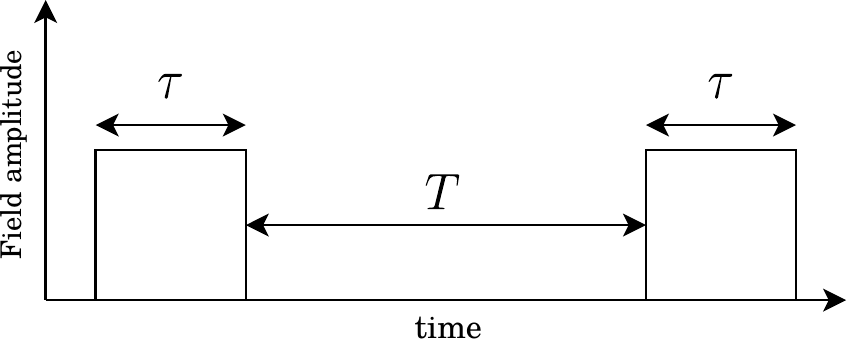}
    \caption{The coherent quantum beat measurement sequence consists of two $\pi$-pulses of width $\tau$ separated by a free-evolution time $T$. The y-axis is the amplitude of the electromagnetic field (either $\Esca$- or $\Bsca$-field, depending on the specific transition).}
    \label{fig:sequence}
\end{figure}

\section{Introduction}
Consider the following problem in precision spectroscopy: an experimenter wants to measure a small energy difference $\omega_e$ between a pair of nearly-degenerate levels $\ket{e_1}$ and $\ket{e_2}$. Directly driving the transition between $e_1 \to e_2$ happens to be impractical due to the smallness of $\omega_e$, and so $\omega_e$ is measured by driving transitions between $\ket{e_1},\ket{e_2}$ and a third state $\ket{g}$. In such a scheme, the broadening of the $g \to e_i$ transitions limits the precision of the $\omega_e$ measurement.

This problem is of particular relevance for experiments that use solid-state samples to search for beyond-Standard-Model physics \cite{Vutha2018,Singh2019,Ramachandran2023}. In these experiments, the putative energy difference $\omega_e$ arises due to the breakdown of some symmetry due to new particles or interactions, and is expected to be very small. Solid-state systems are interesting for precision measurements because of the enormous numbers of particles trapped in their quantum ground state of motion at cryogenic temperatures. There is also a significant practical advantage to be had, since experiments on solid-state samples doped with exotic atoms or isotopes can be simpler to implement than traditional approaches such as laser-cooling and optical trapping. However, solid-state systems are notoriously plagued by inhomogeneous broadening, due to distributions of local crystal fields, impurities or defects. Therefore, a measurement technique to resolve small energy shifts concealed within inhomogeneously-broadened lines would be able to unlock the potential of solid-state systems for precision measurements. Such a technique could also be useful for optical spectroscopy of narrow transitions (e.g., \cite{Thorpe2013}) and for quantum information processing and storage in solid-state systems \cite{Timoney2012,Sellars2015,Zhong2019}.

A well-known method for circumventing inhomogeneous broadening is quantum beat spectroscopy \cite{Aleksandrov1963,Dodd1964,Hadeishi1965}, which has a long history of applications (see, e.g., \cite{Haroche1973,Aleksandrov1979,Vreeker1985,Hack1991,Kawall2004} among others). The essence of this method is to create coherence between the states $\ket{e_1},\ket{e_2}$ using a pulse of light, which leads to interference between the $e_1 \to g$ and $e_2 \to g$ decay paths, causing modulation of the spontaneously emitted fluorescence. The modulation (``beat'') appears at the difference frequency $\omega_e$. However, quantum beat spectroscopy relies on spontaneous emission, which precludes its use in precision measurements of long-lived electronic states, hyperfine states or spin sublevels. 

In this article, we describe a method of coherent quantum beat (CQB) spectroscopy that enables the measurement of small energy differences buried within inhomogeneous broadening, without relying on spontaneous emission. In essence, CQB makes use of interference that arises between the near-degenerate states when coupled by a field to other states. Crucially, we show that this interference is insensitive to the detuning of the field, and thus it is largely immune to inhomogeneous broadening. Therefore the entire inhomogeneously-broadened ensemble can contribute to the measurement, leading to improved precision. We also show that CQB can be extended to four-level systems (two pairs of near-degenerate states), allowing for the measurement of both sets of energy differences, with minimal state preparation. These features of CQB enable precision measurements in hitherto unexplored atomic, molecular and solid-state systems.

\section{Details of the technique} \label{sec:details}
We begin by illustrating the essential steps of the CQB technique in this section, using as an example a simple three-level system. The application of CQB to a more realistic physical system, and extension to four-level systems, will be discussed in Sections \ref{sec:inhomo}, \ref{sec:small Rabi}, \ref{sec:four-level}.

Let the ensemble of atoms participating in the measurement be initially in the state $\ket{g}$ and consider atoms that interact with light that has frequency $\omega$, as shown in the energy level diagram in Fig.\ \ref{fig:3_level_structure}. We denote the average of the $g \to e_1,e_2$ resonance frequencies as $\omega_0$, and define the detuning $\Delta = \omega - \omega_0$. We assume that the Rabi frequency is identical for the $g \to e_i$ ($i=1,2$) transitions, defining it to be $\Omega = \frac{1}{\sqrt{2}}\bra{e_1}H_\mrm{int}\ket{g} = \frac{1}{\sqrt{2}}\bra{e_2}H_\mrm{int}\ket{g}$, where $H_\mrm{int}$ is the atom-field interaction Hamiltonian.

The CQB sequence, shown in Fig.\ \ref{fig:sequence}, consists of two pulses of width $\tau = \pi/\Omega$ at a frequency $\omega=\omega_0$, separated by a variable time interval $T$. The first pulse performs the same role as in traditional quantum beat spectroscopy, creating a coherent superposition of the two states of spectroscopic interest, $\ket{\psi(0)} = \alpha \ket{e_1} + \beta \ket{e_2}$. During the free evolution time $T$ the superposition evolves to $\ket{\psi(T)} = \alpha \ket{e_1} e^{+i \omega_e T/2}+ \beta \ket{e_2} e^{-i \omega_e T/2}$. Since there is no spontaneous emission in this scenario, the second $\pi$-pulse serves to project part of the $e_1,e_2$ superposition back to $\ket{g}$ and closes the quantum-state interferometer. Thus, the system returns to $\ket{g}$ after a free-evolution time $T$ that is an integer multiple of $2 \pi / \omega_e$.
A final measurement of the population in $\ket{g}$ as a function of $T$ yields the energy difference $\omega_e$. The ground state population, $P_g$, can be measured by, e.g., exciting the atoms on an optical transition and measuring the absorption or fluorescence. This interference pattern is the basic building block of CQB.

Quantum-state interferometer sequences similar to the basic scheme described above have been previously used in gas-phase molecular beam measurements (e.g., \cite{Ho2020}). However, it is not immediately evident that such a scheme can still be used when $\omega_0$ is inhomogeneously broadened. 

To understand the CQB sequence in the presence of inhomogeneous broadening, we begin by describing the three-level system using the following Hamiltonian in the $\ket{g}, \ket{e_1}, \ket{e_2}$ basis:
\begin{align}
    H = \begin{pmatrix}
        0 & {\frac{\Omega}{\sqrt{2}} \cos\left(\omega t\right)} & {\frac{\Omega}{\sqrt{2}} \cos\left(\omega t\right)} \\
        {\frac{\Omega}{\sqrt{2}} \cos\left(\omega t\right)} & \omega_0 - \frac{\omega_e}{2} & 0\\
        {\frac{\Omega}{\sqrt{2}} \cos\left(\omega t\right)} & 0 & \omega_0 + \frac{\omega_e}{2} \\
        \end{pmatrix}.
\end{align}
Following the usual rotating-wave approximation, the Hamiltonian becomes
\begin{align*}
\begin{split}
    H &= H_0 + H_{\mathrm{int}} \\ &= \begin{pmatrix}
    0 & 0 & 0 \\
    0 & \Delta - \frac{\omega_e}{2} & 0\\
    0 & 0 & \Delta + \frac{\omega_e}{2}\\
    \end{pmatrix}
    + \begin{pmatrix}
    0 & \frac{\Omega}{2\sqrt{2}} & \frac{\Omega}{2\sqrt{2}} \\
    \frac{\Omega}{2\sqrt{2}} & 0 & 0\\
    \frac{\Omega}{2\sqrt{2}} & 0 & 0\\
    \end{pmatrix}.
\end{split}
\end{align*}

We define the propagators $D_{0} \left(t, t_0\right) = e^{-i H_0 (t-t_0)}$ and $D\left(t, t_0\right) = e^{-i H (t-t_0)}$. An initial state $\ket{\Psi_i}$, after evolving through the CQB sequence, turns into
\begin{align}
    \ket{\Psi_f} = D\left(T+2 \tau, T+\tau\right) D_{0}\left(T+\tau, \tau\right) D\left(\tau, 0\right) \ket{\Psi_i}.
\end{align}
If the system is initially in the state $\ket{\Psi_i} = \ket{g}$, the population $P_g = |\braket{g}{\Psi_f}|^2$ at the end of the sequence can be analytically calculated to be
\begin{widetext}

\begin{align}\label{eq:P_g_long}
\begin{split}
P_g \left(\tau,T,\Delta\right)  = \frac{1}{16}
& \left\{ 8 \sin ^4\left(\frac{\Omega \tau }{2}\right) \cos   \left(\omega_e T\right)+4 \cos (\Omega \tau )+3 \cos (2 \Omega \tau) - 8
\sin ^2(\Omega \tau ) \cos (\Delta  T) \cos \left(\frac{ \omega_e T}{2}\right) +9 \right\} \\
&+ 2\left(\frac{\Delta}{\Omega}\right)  \sin (\tau \Omega) \sin ^2\left(\frac{\Omega \tau}{2}\right) \sin (\Delta  T) \cos \left(\frac{\omega_e T}{2} \right)\\
&+ 2 \left(\frac{\omega _e}{\Omega}\right) \sin \left(\frac{\Omega \tau }{2}\right) \biggl[2 \cos ^2\left(\frac{\Omega \tau }{2}\right) \sin \left(\frac{\omega_e T}{2}\right)
\left(\cos \left(\Delta  T-\frac{\Delta  \tau }{2}\right)-\cos \left(\frac{ 
\Omega \tau}{2}\right) \cos (\Delta  T)\right) \\
&+ \sin ^2\left(\frac{\Omega \tau
}{2}\right) \left(\cos \left(\frac{\Omega \tau }{2}\right)-\cos \left(\frac{\Delta  \tau }{2}\right)\right) \sin \left( \omega_e T\right)\biggr] + \mathcal{O}\left(\frac{\Delta}{\Omega}\right)^2 + \mathcal{O}\left(\frac{\omega_e}{2\Omega}\right)^2.
\end{split}
\end{align}

For $\pi$-pulses, when $\tau=\pi/\Omega$, the above result simplifies to
\begin{align} \label{eq:P_g}
P_g \left(\tau = \frac{\pi}{\Omega}, T, \Delta\right) =
\frac{1}{2} \left\{1 + \cos \left(\omega _e T\right)-4\left(\frac{\omega_e}{\Omega}\right)\cos \left(\frac{\pi  \Delta }{2
   \Omega }\right) \sin \left(\omega _e T\right)\right\},
\end{align}
\end{widetext}
to the same order of approximation as Eq. (\ref{eq:P_g}). As expected, the ground-state population as a function of the pulse length $T$ shows modulations at the desired frequency difference $\omega_e$. We next examine the role of inhomogeneous broadening of the resonance frequency $\omega_0$, or equivalently, the detuning $\Delta$.



\subsection{Three-level system with inhomogeneous broadening} \label{sec:inhomo}
We first present an intuitive sketch of how CQB works despite inhomogeneous broadening. Note that inhomogeneous broadening affects the coherent superposition $\ket{\psi(0)} = \alpha \ket{e_1} + \beta \ket{e_2}$ prepared by the first pulse, because the amplitudes $\alpha$ and $\beta$ depend on the generalized Rabi frequency $\Omega' = \sqrt{\Omega^2 + \Delta^2}$. But the values of $\alpha$ and $\beta$ do not change the phase accumulated during the free-evolution time $T$. The measurement of $\omega_e$ relies only on the phase accumulated, and so it is not significantly affected by the statistical distribution of $\omega_0$ values over the inhomogeneous width $\Gamma_\mrm{inh}$. 


To see this effect quantitatively, we consider a normal distribution of values of $\Delta$ (with mean $\Delta_0$ and full width at half-maximum $\Gamma_\mrm{inh}$) and average $P_g$ from Eq. (\ref{eq:P_g}) over this distribution. The ensemble-averaged ground-state population is
\begin{widetext}
\begin{align} \label{eq:P_g_avg}
\begin{split}
\avg{P_g \left(\tau = \frac{\pi}{\Omega}, T, \Delta\right)} =
\frac{1}{2} \Biggl\{&1 + \cos \left(\omega _e T\right)-4\left(\frac{\omega_e}{\Omega}\right) \cos \left(\frac{\pi  \Delta_0 }{2
   \Omega }\right) \exp\left[-\frac{\Gamma_\mrm{inh}^2}{16 \ln 2} \left(\frac{\pi}{2\Omega}\right)^2\right] \sin \left(\omega _e T\right) \Biggr\}.
\end{split}
\end{align}
\end{widetext}
We find that the fringes in $\avg{P_g}$ are largely independent of $\Delta_0$ and $\Gamma_\mrm{inh}$. The parameters of the inhomogeneous distribution affect the amplitude of the $\sin(\omega_e T)$ term in Eq.\ (\ref{eq:P_g_avg}), which results in an overall fringe shift but \emph{does not change} the periodicity of the oscillations as a function of $T$. Therefore, $\omega_e$ can be extracted by measuring the dependence of $\avg{P_g}$ on $T$. Importantly, the energy difference $\omega_e$ can be cleanly measured despite overwhelmingly larger inhomogeneous broadening, $\Gamma_\mrm{inh} \gg \omega_e$.


\begin{figure}[h!]
    \centering
    \includegraphics[width=\columnwidth]{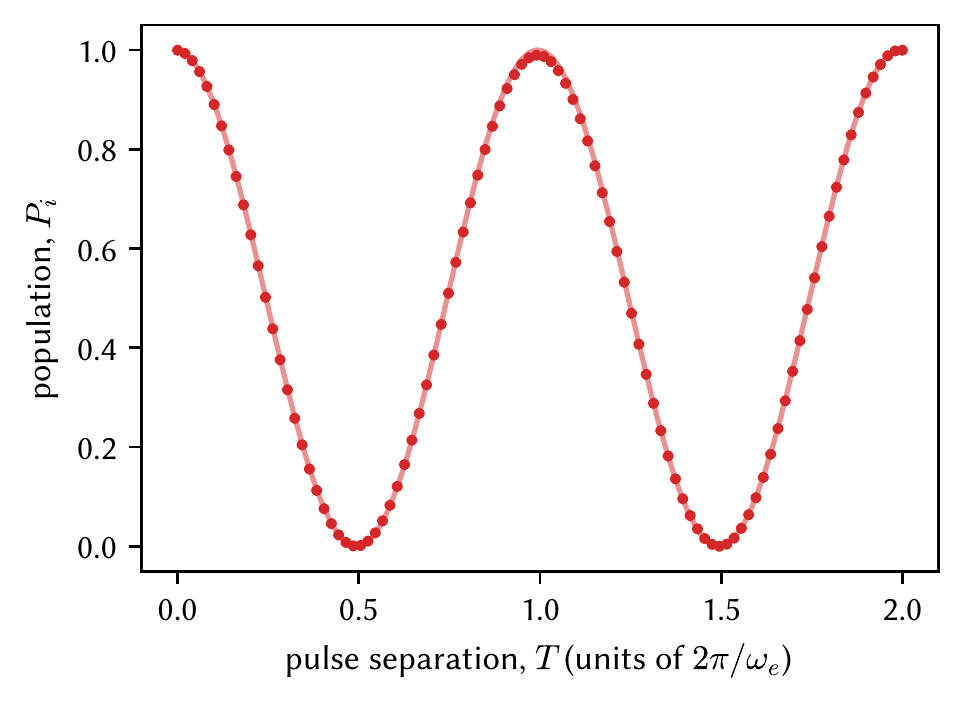}
    \caption{Population in the ground state (points) after a resonant CQB sequence for different free-evolution times, in the presence of inhomogeneous broadening of the $g \to e_i$ transitions. The simulation parameters were $\omega_0 = 1$, $\Omega = 10^{-2}$ and $\omega_e = 10^{-4}$. The population traces were averaged over $N=100$ samples for the detuning $\Delta$ drawn from a normal distribution with zero mean and width $\Gamma_\mrm{inh} = 1 \times 10^{-3}$. The solid line through the points is the analytic expression shown in Eq.\ (\ref{eq:P_g_avg}).}
    \label{fig:3_level_inhom_avg}
\end{figure}

This feature of CQB can be confirmed using numerical simulations without approximations. Fig.\ \ref{fig:3_level_inhom_avg} shows the results of a CQB sequence applied to an ensemble of three-level atoms whose $\omega_0$ values were drawn from a normal distribution with width $\Gamma_\mrm{inh}$. There is excellent agreement with our analytical model, showing that coherent quantum beats in the ensemble-averaged ground-state population persist despite inhomogeneous broadening. 

\subsection{Pulse imperfections}

We now examine the sensitivity of the CQB technique to pulse width and pulse height errors. To analyze these effects, we take a first-order series expansion of Eq. (\ref{eq:P_g_long}) around both $\tau = \frac{\pi}{\Omega}$ and $\Delta = 0$:
\begin{widetext}
\begin{align} \label{eq:pulse_error}
\begin{split}
    P_g \left(\tau \sim \frac{\pi}{\Omega}, T, \Delta \sim 0 \right) = \frac{1}{2} \left(\cos\left(\omega_e T\right)+1\right) +
    \left(\frac{\omega_e}{\Omega}\right) (-\Omega \tau +\pi -2) \sin \left(\omega_e T\right) + \mathcal{O}\left(\tau - \frac{\pi}{\Omega}\right)^2 + \mathcal{O}\left(\Delta\right)^2.
\end{split}
\end{align}
\end{widetext}
Both the terms in Eq.\ (\ref{eq:pulse_error}) display modulations at $\omega_e$ as a function of $T$, as expected. The second term depends on $\left(\tau - \frac{\pi}{\Omega}\right)$, but (a) this term is suppressed by the small number $\omega_e / \Omega$, and (b) this term only leads to a shift of the fringes, but does \emph{not} affect their periodicity versus $T$. Therefore the measurement of $\omega_e$ is insensitive to pulse timing errors to first order. We also note that there is no first-order dependence on $\Delta$.

\begin{figure}
    \centering
    \includegraphics[width=\columnwidth]{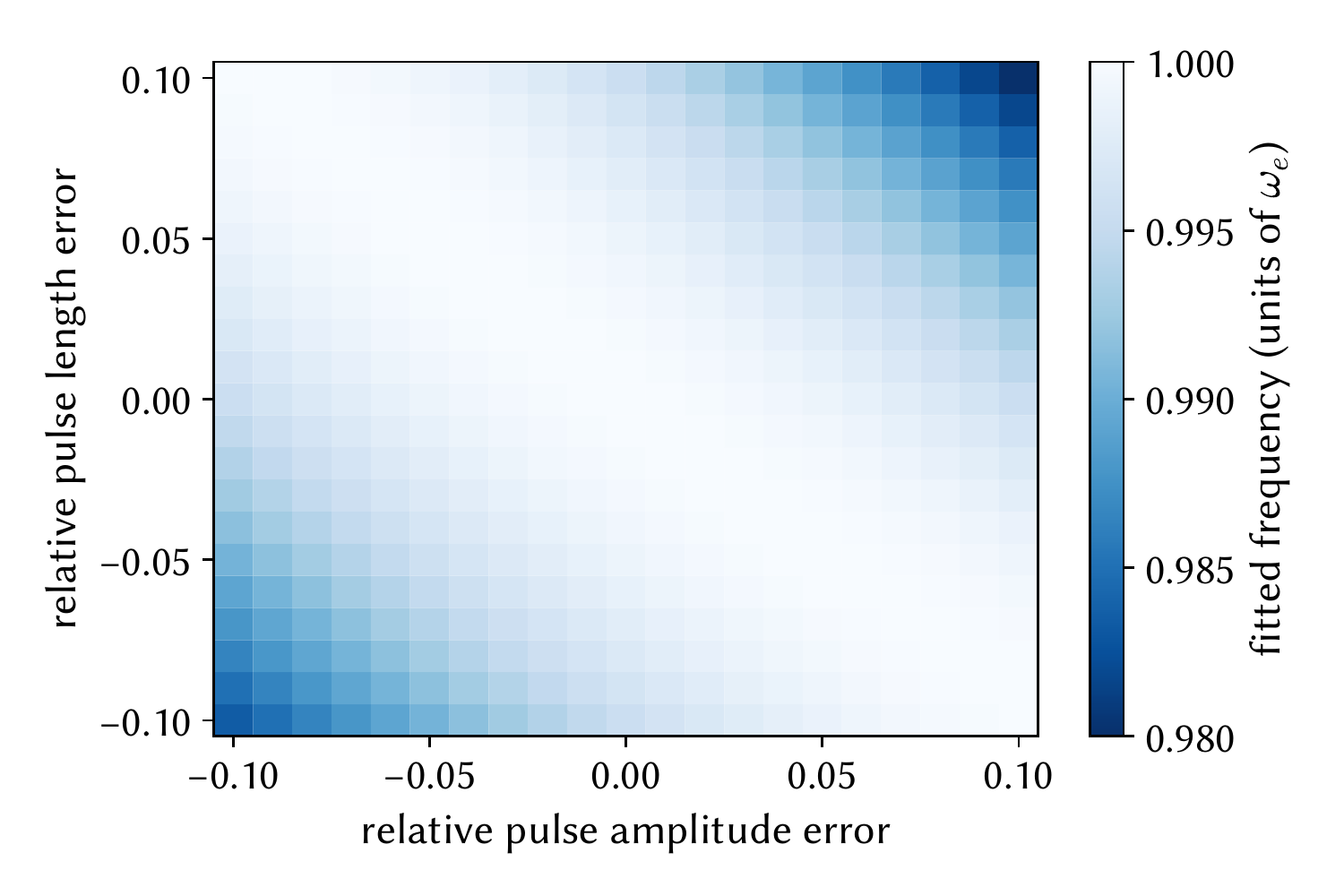}
    \caption{Modulation frequency of the ground state population after a resonant CQB sequence, as the pulse length and amplitude are changed from perfect $\pi$-pulses. The simulation parameters were $\omega_0 = 1$, $\omega_e = 10^{-4}$, $\Omega = 10^{-2}$, and $\Delta = 0$.}
    \label{fig:frequency_error}
\end{figure}


We use numerical simulations to understand higher-order effects of pulse imperfections and evaluate the robustness of CQB. We consider the effect of  $\pm 10$\% errors in the pulse length from $\pi/\Omega$, and $\pm 10\%$ errors in the Rabi frequency $\Omega$ compared to $\pi/\tau$, on the size and periodicity of the coherent quantum beats. We fit the numerically-calculated ground state population to a sum of two sinusoids of variable frequencies, amplitudes, and phases. 
Figs.\ \ref{fig:frequency_error} shows that the modulation frequency is only affected to second order in the properties of the pulse. 

Therefore CQB is practically useful as a {precision} spectroscopy technique to measure small frequency shifts that are masked by inhomogeneous broadening. We anticipate that a standard spectroscopy technique (e.g., Ramsey interferometry) would be used to estimate $\omega_0$, and after adjusting $\Delta \approx 0$ and $\Omega \tau \approx \pi$, a CQB measurement can be used to obtain the splitting $\omega_e$ to high precision. 

\begin{figure}
    \centering
    \includegraphics[width=\columnwidth]{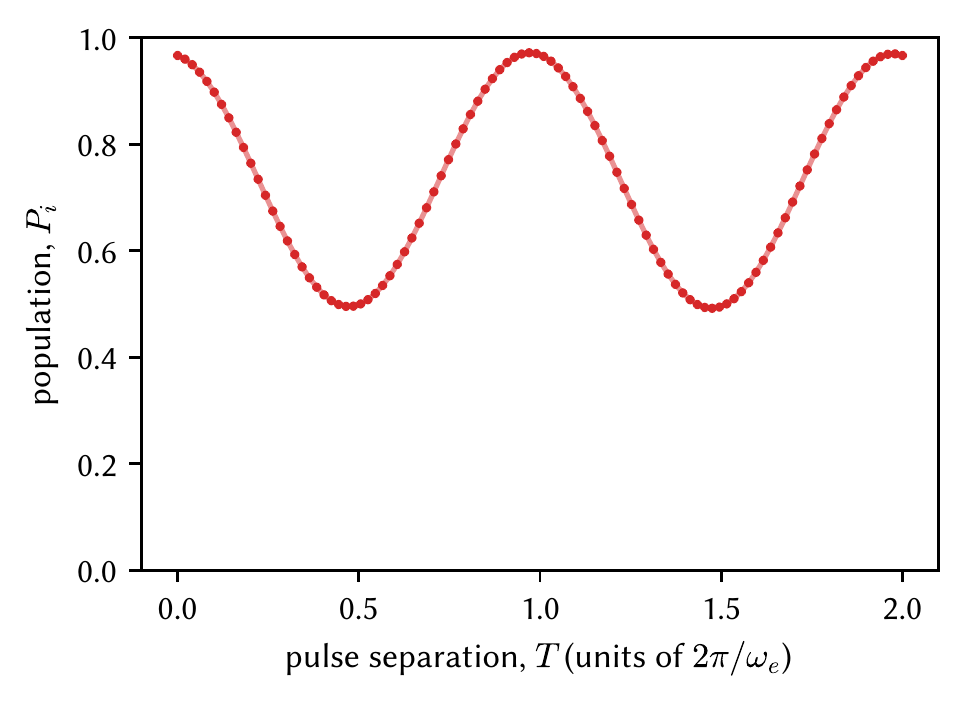}
    \caption{Population in the ground state after a CQB sequence with rapid adiabatic passage for different free-evolution times, in the presence of inhomogeneous broadening of the $g \to e_i$ transitions. The simulation parameters were $\omega_0 = 1$, $\Omega = 10^{-3}$, $\omega_e = 10^{-4}$, $R=20$, and $S=80$ (parameters described in the text). The population traces were averaged over $N=100$ samples for the detuning $\Delta$, drawn from a normal distribution with zero mean and width $\Gamma_\mrm{inh} = 3 \times 10^{-3}$. The solid line through the points is a guide to the eye.}
    \label{fig:3_level_rap}
\end{figure}

\subsection{CQB with rapid adiabatic passage} \label{sec:small Rabi}
In the above discussion, we assumed for analytical ease that the detuning $\Delta$ (and therefore also the broadening $\Gamma_\mrm{inh}$) was small compared to the Rabi frequency $\Omega$. Here we explore the consequences of relaxing this assumption. This regime corresponds to ``weak'' excitation, where the power broadening is much smaller than the inhomogeneous broadening.  

One obvious consequence is that a field pulse with $\Omega \ll \Gamma_\mrm{inh}$ can only interrogate a small fraction of the inhomogeneously-broadened ensemble, which would result in a significant loss of signal. But remarkably, it is possible to make CQB measurements even in the $\Omega \ll \Gamma_\mrm{inh}$ limit without incurring any penalties of loss of signal.
The key idea is that, instead of using $\pi$-pulses to transfer population from $\ket{g}$ to the coherent superposition of $\ket{e_1}, \ket{e_2}$, lower-power rapid adiabatic passage (RAP) sweeps can be used to accomplish the same task. The modified CQB sequence then begins with a RAP field pulse $F(t)$ whose frequency is swept over a range $\Gamma_\mrm{inh} \equiv R \Omega > \Gamma_\mrm{inh}$ centered on $\omega_0$ during a time interval $\tau_{\mathrm{RAP}} = S \tau = S \pi / \Omega$. Fixing the chirp rate $\alpha = \Gamma_\mrm{inh} / \tau_{\mathrm{RAP}}$, we can pick $R$ and $S$ to satisfy the RAP criteria: $|\alpha| / \Omega^2 < 1$ and $|\alpha| \tau_{\mathrm{RAP}}^2 \gg 1$ \cite{Malinovsky2001}. In this adiabatic version of the CQB sequence, the first RAP pulse is followed by a free evolution time $T$ and a conjugate time-reversed RAP pulse $\bar{F}(t)$ that returns the population to $\ket{g}$. The numerical results shown in Fig.\ \ref{fig:3_level_rap} confirm that well-defined oscillations in the ensemble-averaged ground-state population are obtained at the end of the sequence, allowing $\omega_e$ to be measured exactly as before. Therefore the CQB technique offers a way to measure small energy differences masked by significantly larger amounts of inhomogeneous broadening, across a range of parameter regimes.

\subsection{Four-level systems with incoherent mixtures} \label{sec:four-level}
\begin{figure}
    \centering
    \includegraphics[width=0.75\columnwidth]{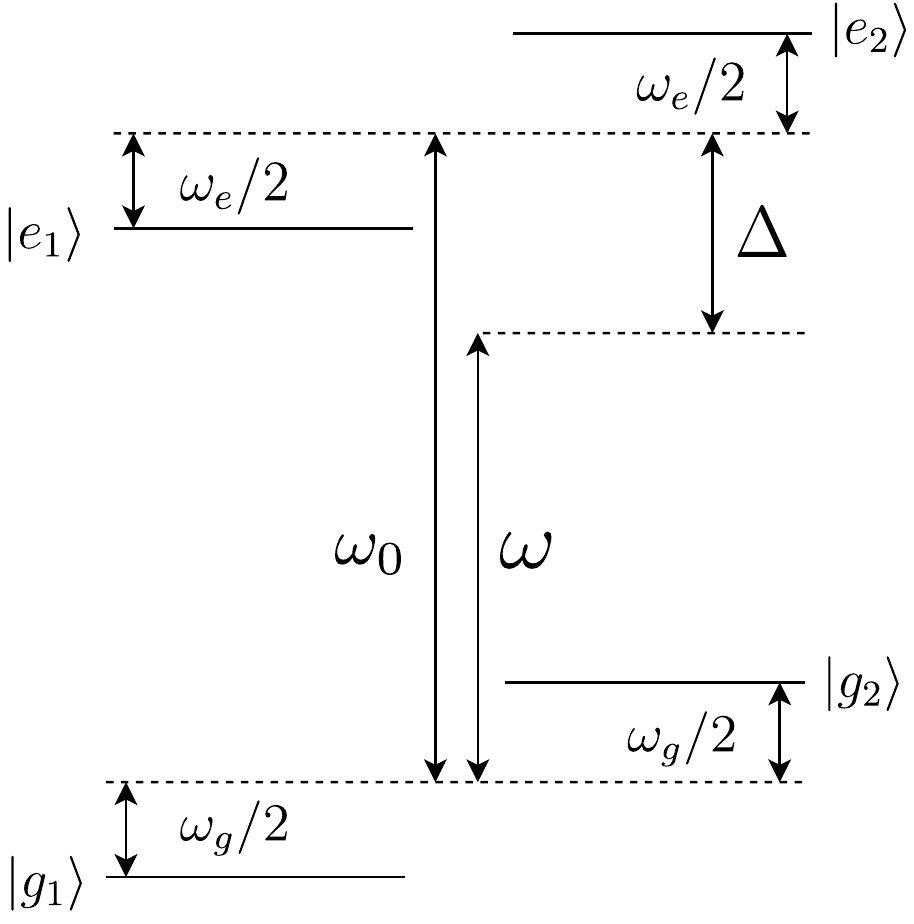}
    \caption{A system of four levels, where the energy differences of interest ($\omega_g, \omega_e$) are measured using spectroscopy of the $g_i \to e_j$ transitions $ (i,j=1,2)$. The frequency of light is $\omega$ and the average of the $g_i \to e_j$ resonance frequencies is $\omega_0$.} 
    \label{fig:4_level_structure}
\end{figure}
The foregoing discussion of the CQB technique has been based on the three-level V-type system shown in Fig.\ \ref{fig:3_level_structure}. But there are a number of experimentally relevant situations, especially in the context of precision measurements in solid-state systems (see e.g., \cite{Singh2019,Li2022, Ramachandran2023}), where more levels may be present.

Therefore we examine whether a similar technique can be made to work in a four-level system. Consider the system of four states in Fig.\ \ref{fig:4_level_structure}, where the two ground states $\{\ket{g_1}, \ket{g_2}\}$ are separated by $\omega_g$ and the two excited states $\{\ket{e_1}, \ket{e_2}\}$ are separated by $\omega_e$. We assume that $\omega_g,\omega_e \ll \omega_0$. In solid-state systems, such a level structure can arise in systems with spin $F \geq 3/2$ in the presence of quadrupole interactions or crystal fields \cite{Longdell2006}. In the context of fundamental symmetry violation searches, $\{\ket{g_1}, \ket{g_2}\}$ and $\{\ket{e_1}, \ket{e_2}\}$ can be Kramers doublets in a half-integer-spin system, where the small energy differences $\omega_g, \omega_e$ are produced by violation of time-reversal symmetry \cite{Ramachandran2023}. 

\begin{figure*}
    \centering
    \includegraphics[width=\textwidth]{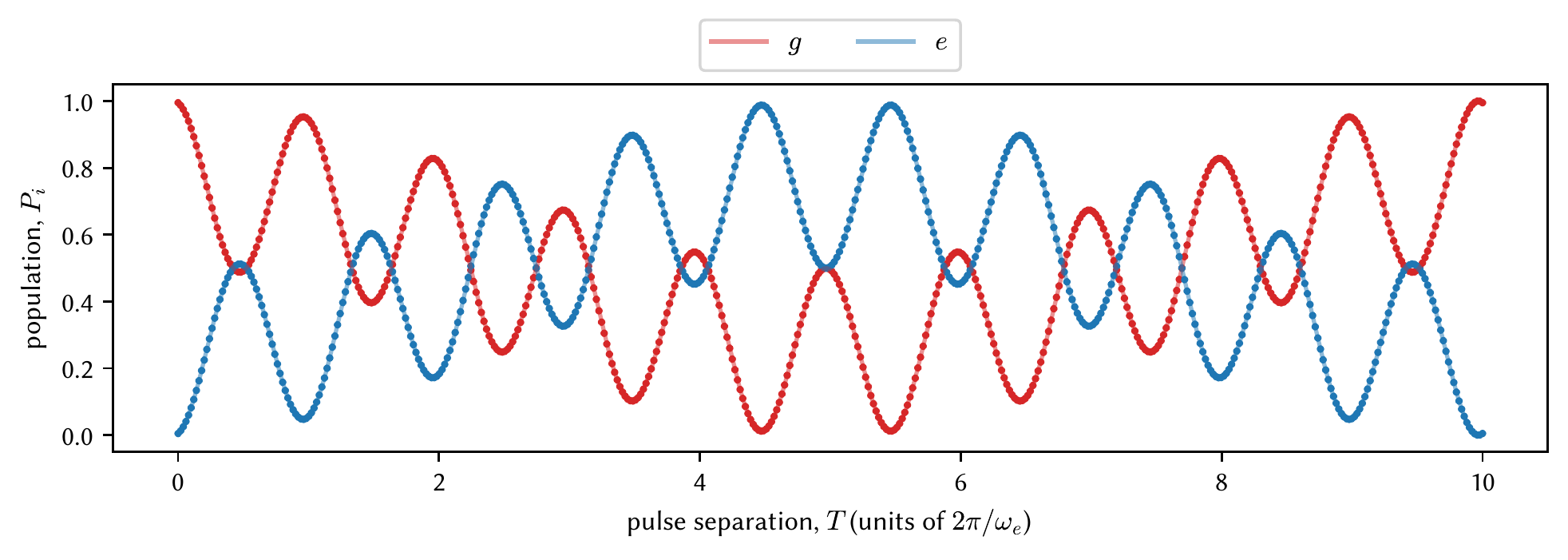}
    \caption{Populations in a four-level system after a resonant CQB sequence for varying free-evolution times. $g$ ($e$) denote the summed populations of both ground (excited) states. The simulation parameters were $\omega_0=1$, $\Omega = 10^{-2}$, $\omega_e = 10^{-3}$, $\omega_g = 10^{-4}$, and $\Delta = 0$.}
    \label{fig:4_level_resonant}
\end{figure*}
To model a typical situation in such solid-state systems, we assume that the ground states $\ket{g_1}$ and $\ket{g_2}$ are initially populated and are in thermal equilibrium at a temperature $T_\mrm{spin} \gg \omega_g/k_B$, so that the density matrix is initially in the mixed state $\rho(0) = 1/2 \ket{g_1}\bra{g_1} + 1/2 \ket{g_2}\bra{g_2}$. To understand the effect of the CQB sequence on such mixed states, we numerically evolve the density matrix $\rho$ of the four-level system through the CQB pulse sequence, with the four-level Hamiltonian
\begin{align}
    H = \begin{pmatrix}
        -\frac{\omega_g}{2} & 0 & {\frac{\Omega}{2} \cos\left(\omega t\right)} & {\frac{\Omega}{2} \cos\left(\omega t\right)} \\
        0 & +\frac{\omega_g}{2} & {\frac{\Omega}{2} \cos\left(\omega t\right)} & {\frac{\Omega}{2} \cos\left(\omega t\right)} \\
        {\frac{\Omega}{2} \cos\left(\omega t\right)} & {\frac{\Omega}{2} \cos\left(\omega t\right)} & \omega_0 - \frac{\omega_e}{2} & 0\\
        {\frac{\Omega}{2} \cos\left(\omega t\right)} & {\frac{\Omega}{2} \cos\left(\omega t\right)} & 0 & \omega_0 + \frac{\omega_e}{2} \\
        \end{pmatrix}.
\end{align}
Here we assume that the pulses couple each ground state to the two excited states with equal matrix elements, so that $\Omega = \frac{1}{2} \bra{e_i}H_\mrm{int}\ket{g_j}$ ($i,j = 1,2$). This is a reasonable assumption if $\{\ket{g_1}, \ket{g_2}\}$ and $\{\ket{e_1}, \ket{e_2}\}$ are Kramers pairs with a small energy difference from a violation of time-reversal symmetry. 

\begin{figure}
    \centering
    \includegraphics[width=\columnwidth]{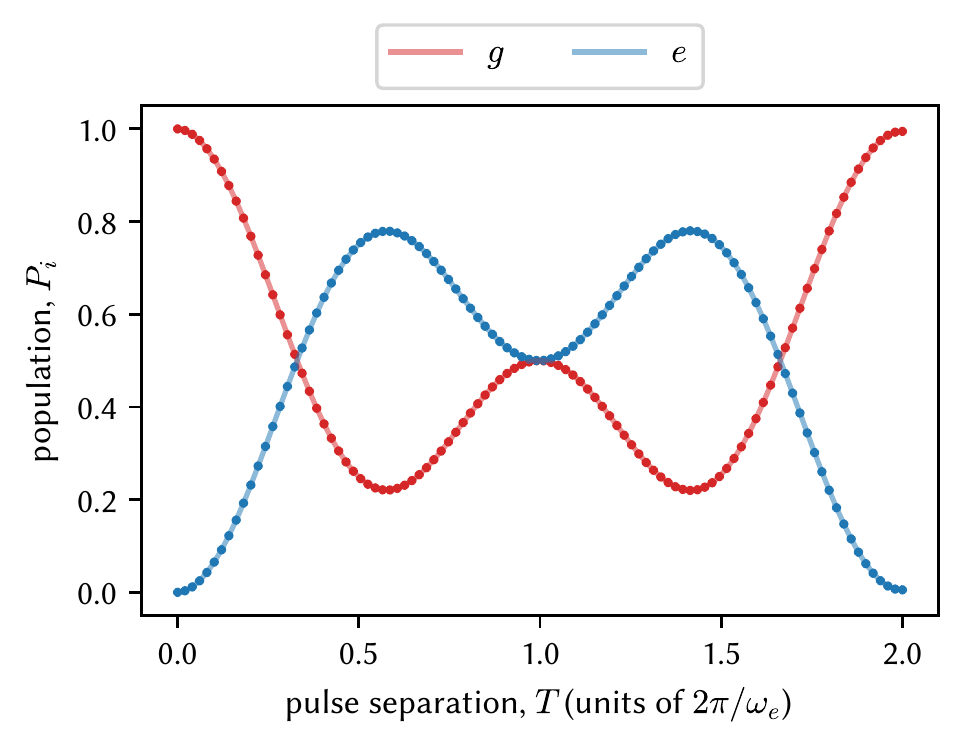}
    \caption{Populations in a four-level system after a resonant CQB sequence for varying free-evolution times, in the presence of inhomogeneous broadening of the $g_i \to e_j$ transitions. The simulation parameters were $\omega_0 = 1$, $\Omega = 10^{-2}$, $\omega_e = 10^{-4}$, and $\omega_g = 5 \times 10^{-5}$. A set of $N=100$ samples of the detuning $\Delta$ were drawn from a zero mean normal distribution with width $\Gamma_\mrm{inh} = 2.5 \times 10^{-3}$. }
    \label{fig:4_level_inhom}
\end{figure}

Fig.\ \ref{fig:4_level_resonant} shows the effect of the CQB sequence on a four-level atom when $\Delta=0$. Despite starting from the mixed state $\rho\left(0\right)$, the pattern of fringes in the ground state population at the end of the sequence shows beats at both $\omega_e$ and $\omega_g$, allowing both energy differences to be measured. This elucidates a more general and surprising feature: the CQB method can work in four-level systems without requiring the ground states to be prepared in coherent superpositions. This feature greatly simplifies the measurement process, leading to experimental simplicity and improvements to the duty cycle of measurements. We further verified that CQB works when inhomogeneous broadening of the $g_i \to e_j$ transitions is included, in the regime $\Omega > \Gamma_\mrm{inh} > \max\{\omega_e, \omega_g\}$. The calculations in Fig.\ \ref{fig:4_level_inhom} for a distribution of resonance frequencies demonstrate that clear fringes in $\avg{P_g}$ are obtained despite broadening, and that the \textit{entire ensemble} contributes to the measurement of the small energy differences $\omega_g,\omega_e$.

\section{Discussion}

\subsection{Sensitivity}
The CQB technique amounts to mapping out an $\omega_e$ and/or $\omega_g$ fringe by varying $T$, while keeping $\tau$ and $\Delta$ fixed, and then fitting a sinusoid to the periodic modulation of the ground-state population. This procedure may seem qualitatively different from conventional spectroscopy techniques such as Ramsey interferometry, where the field detuning is usually the parameter that is varied. Nonetheless, the two techniques have comparable fundamental limits to their precision. For CQB the attainable precision is \cite{Montgomery1999}
\begin{align}\label{eq:CQB_sensitivity}
    \delta \omega_{\mathrm{CQB}} / 2\pi = \frac{1}{\pi \mathcal{S} } \sqrt{\frac{6}{T_{\mathrm{int}} T}},
\end{align}
where $T_{\mathrm{int}}$ is the total integration time and $\mathcal{S}$ is the signal-to-noise ratio of a measurement. 
Here we have assumed that a measurement is made once every pulse separation time $T$ (i.e. $T \gg \tau$). In contrast, the Ramsey spectroscopy technique is capable of precision
\begin{align}\label{eq:Ramsey_sensitivity}
    \delta \omega_{\mathrm{Ramsey}} / 2\pi = \frac{1}{2 \mathcal{S}} \sqrt{\frac{1}{T_{\mathrm{int}} T'}},
\end{align}
where $T'$ is the separation between the oscillatory pulses in the Ramsey sequence.

There are two important sensitivity advantages to the CQB technique, as evident from Eq. (\ref{eq:CQB_sensitivity}) and (\ref{eq:Ramsey_sensitivity}). First, note that $\delta \omega$ in both cases improves $\propto 1/\mathcal{S}$. To measure an energy difference masked by inhomogeneous broadening using Ramsey interferometry, the linewidth must be narrowed in some way so that the $g \to e_i$ transitions are distinguishable by the state readout transition (with inhomogeneous width $\Gamma^{\prime}_\mrm{inh}$). In solid-state systems, techniques such as spectral hole-burning \cite{Abragam2012} can yield lines narrower than the inhomogeneous width, but always at the expense of signal-to-noise because only a fraction $\sim \frac{2 \omega_e}{\Gamma^{\prime}_\mrm{inh}}$ of the atoms in the ensemble are selected.
Meanwhile, \emph{all} the atoms in the ensemble contribute to the CQB signal, resulting in significantly higher signal-to-noise $\mathcal{S}$ in precision measurements. 

Second, in order to minimize $\delta \omega$, the free evolution time ($T$ in CQB, and $T'$ in Ramsey) must be maximized. The free evolution time can be increased without signal loss up to the ensemble phase coherence time. In Ramsey spectroscopy, the ensemble coherence time $T_2 \sim 1/\Gamma_\mrm{inh}$ is limited by the broadening of the $g_i \to e_j$ transition frequency $\omega_0$.  
In contrast, the CQB method is insensitive to $\Gamma_\mrm{inh}$. The evolution time $T$ is limited only by the coherence time $T_{c}$ of the superposition of states $g_i$ or $e_j$, not the $g_i - e_j$ superpositions. Therefore the spectroscopic resolution improves as $\delta \omega \sim 1/\sqrt{T_{c}}$ rather than $\sqrt{\Gamma_\mrm{inh}}$. Taken together, these two advantages can lead to significant improvements, by a factor $\mathcal{X} = \pi \sqrt{\frac{\Gamma_\mrm{inh} \Gamma^{\prime}_\mrm{inh}  T_c}{12 \omega_e}}$, compared to the Ramsey method.

\subsection{A specific example: Eu$^{3+}$:Y$_2$SiO$_5$}



The advantages of the CQB technique discussed in the preceding section make it especially well-matched to precision solid-state spectroscopy because (a) $\Gamma_\mrm{inh}$ and $\Gamma^{\prime}_\mrm{inh}$ can be large, and (b) $T_c$ can be much longer than the ensemble coherence time $T_2 \sim 1/\Gamma_\mrm{inh}$. As a specific example, we consider Eu$^{3+}$:Y$_2$SiO$_5$. This system is well-studied in the context of quantum information storage \cite{Macfarlane2004, Longdell2006, Timoney2012, Zhong2015}, and has been proposed as a platform to search for new T-violating physics \cite{Ramachandran2023}. The T-violation search relies on the measurement of a small energy difference between nominally degenerate pairs of states (``Kramers doublets'').

The relevant states in Eu$^{3+}$:Y$_2$SiO$_5$ have the structure of the four-level system in Fig.\ \ref{fig:4_level_structure}. The $g_i \to e_j$ hyperfine transition is  
inhomogeneously broadened ($\Gamma_\mrm{inh}^{\mathrm{hfs}} \sim 70$ kHz \cite{Timoney2012}), presenting a challenge to the measurement of sub-mHz separations between the $e_j$ states. However, the hyperfine states have notably long coherence times ($T_{c}^{\mathrm{hfs}} \sim 15$ ms) as measured through hole-burning spectroscopy \cite{Alexander2007}. Application of the CQB method to this system therefore enables precision measurement of the small energy differences between Kramers doublets.

From a free evolution time of $T = T_{c}^{\mathrm{hfs}} \approx 15$ ms with CQB as compared to $T' = T_2 \sim 1/\Gamma_\mrm{inh}^{\mathrm{hfs}} \approx 2\,\mu$s with Ramsey, there is a $\sim 55 \times$ improvement in precision. We further estimate that $10^4 \times$ more atoms from the ensemble can be used in the CQB measurement by relaxing the need for spectral holeburning \cite{Zhong2015} for state readout, leading to an overall improvement in $\delta \omega$ by over 3 orders of magnitude.



\section{Summary}
We have described a coherent quantum beat spectroscopy method to measure small energy differences even when they are masked by significant inhomogeneous broadening. Whereas established spectroscopy techniques for solid-state systems have to either contend with limited coherence time due to inhomogeneous broadening and/or reduced signal-to-noise by selecting only a small fraction of the ensemble, the CQB method described in this paper takes full advantage of the coherence between the states of spectroscopic interest and uses all the atoms in the ensemble. We have shown that the method is suitable for spectroscopy of multi-level systems, and can take advantage of adiabatic passage to address the full ensemble. We anticipate that this method will lead to useful improvements to the precision of measurements in solid-state systems.

\bibliography{cqb}

\end{document}